\documentclass[11pt,a4paper]{article}
\usepackage{jheppub}
\usepackage{amsmath}
\usepackage{amssymb}

\newcommand{\bea}{\begin{eqnarray}}
\newcommand{\eea}{\end{eqnarray}}
\newcommand{\bit}{\begin{itemize}}
\newcommand{\eit}{\end{itemize}}

\def\nl{\nonumber \\}

\def\a{\alpha}

\def\b{\beta}

\def\s{\sigma}

\def\p{\partial}

\def\li{\mathcal{L}}

\def\le{\left(}
\def\ri{\right)}

\def\beq{\begin{equation}}
\def\eeq{\end{equation}}

\def\arr{{\rightarrow}}

\def\pa{{\, \, \, \, \,}}
\def\commuA{(\s_{[1} D_A \s_{2]})}
 
\def\commuC{(\s_{[1}D_A D^2 \s_{2]} )}

\title{On Newton-Cartan trace anomalies} 

\author[a,b]{Roberto Auzzi,} 
\author[a]{Stefano Baiguera} 
\author[a,c]{and Giuseppe Nardelli}

\affiliation[a]{Dipartimento di Matematica e Fisica, Universit\`a Cattolica
del Sacro Cuore, \\
Via Musei 41, 25121 Brescia, Italy}
\affiliation[b]{INFN Sezione di Perugia, \\ Via A. Pascoli, 06123 Perugia, Italy}
\affiliation[c]{TIFPA - INFN, c/o Dipartimento di Fisica, Universit\`a di Trento, 38123 Povo (TN), Italy}

\emailAdd{roberto.auzzi@unicatt.it}
\emailAdd{giuseppe.nardelli@unicatt.it}
\emailAdd{stefano.baiguera92@gmail.com}

\abstract{
We classify the trace anomaly for parity-invariant non-relativistic
Schr\"odinger theories in $2+1$ dimensions coupled to
 background Newton-Cartan gravity.
The general anomaly structure looks very different from the one
 in the $z=2$ Lifshitz theories.  
 The type A content of the anomaly is remarkably identical to that of
  the relativistic $3+1$ dimensional case, suggesting the conjecture that an $a$-theorem
   should exist also in the Newton-Cartan context.

Erratum: due to an overcounting of the number of linearly-independent
terms in the basis, the type A anomaly disappears if Frobenius condition is imposed. 
See appended erratum for details. 
This crucial mistake was pointed out to us in  arXiv:1601.06795. }


\begin{document}

\maketitle


\begin{center}
  \begin{minipage}{0.8\textwidth}
    \vspace{-10pt}
The original claim of the paper, that a type A anomaly exists
if Frobenius condition is imposed, turns out to be false due
to an overcounting of the number of linearly-independent
terms in the basis. The work in \cite{Arav:2016xjc} 
correctly pointed out this mistake; we agree with their results.
See appended erratum for details.
  \end{minipage}
\end{center}

 \section{Introduction}
\label{sec_intro}

Renormalization group (RG) can be understood as a set of trajectories
 in the space of theories
from the UV  to IR; the implicit coarse graining procedure  
suggest that the number of effective degrees of freedom  
decreases from UV to IR. This intuition is correct just if a correct 
counting of effective degrees of freedom is used.
In the case of relativistic $1+1$ unitary dimensional theories, 
it exists indeed a function, introduced by Zamolodchikov \cite{Zamolodchikov:1986gt},
which is monotonically decreasing along a RG trajectory.
This function coincides with the central charge $c$ of the conformal field theory (CFT)
in the UV and in the IR, which is related to the trace
anomaly of the energy-momentum tensor:
\beq
T^\mu_\mu = \mathcal{A}=c R \, .
\eeq

Some of these results can be extended to four dimensional theories.
In particular it was conjectured in \cite{Cardy:1988cwa}
that such monotonically decreasing function exists and that it coincides
with the conformal anomaly coefficient $a$ at the conformal fixed point\footnote{
In this equation the anomaly coefficient $a'$ is a scheme-dependent quantity \cite{Bonora:1985cq},
while $a$ and $c$ are genuine scheme-independent anomalies.
Conformal anomalies have a long history, see e.g. \cite{Duff:1977ay,Duff:1993wm}.}:
\beq
T^\mu_\mu = \mathcal{A}= a E_4 - c W_{\mu \nu \a \b}^2 + a'D^2 R\, .
\eeq
A perturbative proof of this conjecture was given in 
\cite{Osborn:1989td,Jack:1990eb,Osborn:1991gm}.
A non-perturbative proof of a weak version of this theorem
(which states that the central charge $a$ of the IR conformal
fixed point is less than the one in the UV, if both these
fixed points exist) was  given in \cite{Komargodski:2011vj,Komargodski:2011xv}.

In odd dimensions there is no relativistic trace anomaly. Still, there is
a candidate for a monotonically decreasing function of the coupling
along the RG flow, which is related to a scheme-independent
contribution to the free-energy of the theory on a sphere
 \cite{Myers:2010tj,Jafferis:2011zi}.
This suggests that for relativistic theories the RG flow is an irreversible
process as higher momentum modes are integrated out.
In even dimension,
the central charges $c$ and $a$ provide a definition of the effective
number of degrees of freedom that is monotonically decreasing from UV to IR.

It is interesting to ask if similar quantities may exist in the case of 
non-relativistic RG flows. Non-relativistic
scale invariance is realized in Nature in interesting
condensed matter systems, such as unitary Fermi gases.
Both in the Lifshitz and in the Galilean case, scale invariance is in general
characterized by a different scaling
of the time and space coordinates, which can be parameterized by the dynamical
exponent $z$:
\beq
x^i \arr e^{\s} x^i \, ,\qquad t \arr e^{z \s} t \, . 
\eeq
Conformal invariance 
is achieved just in the Galilean case
for $z=2$ (Schr\"odinger invariance)
and in the relativistic case ($z=1$).

In the case of Lifshitz theories,
a detailed study of trace anomalies for various dimensions
and values of $z$ was 
carried on in \cite{Adam:2009gq,Baggio:2011ha,Griffin:2011xs,Arav:2014goa}. 
The result does not give any reasonable
candidate for a decreasing $a$-function; several anomalies are indeed
possible at the scale-invariant fixed points, but their Weyl variation vanishes
identically (type B anomalies \cite{Deser:1993yx}). An analysis like the one
developed  in \cite{Osborn:1989td,Jack:1990eb,Osborn:1991gm} for relativistic 
theories would suggest that no 
monotonically-decreasing anomaly coefficient is present
in the Lifshitz case.

The analysis of the Newton-Cartan (NC) conformal
 anomaly was initiated in \cite{Jensen:2014hqa},
 where an infinite number
of possible terms entering the anomaly was found. In this situation, it is difficult to
figure out
the existence of an $a$-theorem, due to the infinitely many coefficients that are in principle present, and the infinite number of Wess-Zumino consistency conditions to solve.
With these premises, the natural conclusion would be that non-relativistic theories can not admit
an $a$-theorem: either there are not type A anomalies (Lifshitz theories) or there are too many (Schr\"odinger theories).  We shall show that this is not the case.

We shall perform an analysis of the conformal anomalies for Schr\"odinger theories in $2+1$ dimensions.  The discussion of these systems is not moot,
 because there are several physical models which realize these symmetries, 
the simplest example being free fields.
Conformal invariance of $2+1$-dimensional systems is relevant, for example, in anyons
\cite{Bergman:1993kq}. In addition, investigation of $2+1$ dimensional non-relativistic
CFTs gives useful insights for fermions at unitarity in $3+1$ dimensions \cite{Nishida:2007pj}.
Other interesting examples are provided by holography, e.g.~\cite{Son:2008ye,Balasubramanian:2008dm}.

The analysis of the anomaly is important in order 
to identify possible quantities which may decrease along the non-relativistic RG flow.
For this purpose, the non-relativistic theory is coupled to a NC background with torsion.
As a tool to write terms which respect all the symmetries of the problem, we shall use 
discrete light-cone quantization (DLCQ): the NC background
is obtained via dimensional reduction of a relativistic theory on a null circle \cite{Duval:1983pb,Duval:1984cj}.
We shall show that the requirement that the NC background respects causality, 
drastically reduces the possible terms in the anomaly to a finite number. 
We classify them and we solve the Wess-Zumino consistency conditions
in the case of parity-invariant theories. 
We find that 
\beq
2 T^0_0+T^i_i=\mathcal{A}= a \s E_4 - c \s W^2 + b  \s J^2  + \mathcal{A}_{\rm ct} \, ,
\eeq
where the $\mathcal{A}_{\rm ct}$ (see eq.~(\ref{schemedependent})) 
corresponds to terms which can be eliminated
by scheme-dependent counterterms.
The terms $E_4$ and $W^2$ are the DLCQ reduction of the four-dimensional
Euler density and squared Weyl tensor; the term $J^2$ is defined in eq.~(\ref{jj}).
Both $W^2$ and $J^2$ have vanishing Weyl variation
(type B anomalies).
Instead, there is only one term  which has a non vanishing Weyl variation and,
remarkably, it is precisely the same term of the corresponding $3+1$ dimensional theory. It is the
$E_4$ term, and its coefficient is the natural candidate for a monotonically decreasing function from the UV to the IR.

In $d$ spatial dimensions\footnote{In $d=0$ the anomaly vanishes because no 
 non-zero scalar can be built from the NC
geometrical data.}, the NC
anomaly can appear only for discrete values of $z$.
 More precisely (see appendix \ref{AppeA}), for
 \beq z=2n-d \, , 
 \label{discre}
 \eeq
 where $n$ is a positive integer.
It is possible to extend the analysis to other values of $d$ and $z$,
 satisfying eq.~(\ref{discre}).
 The Weyl weight of $E_4$
is $-4$ for every $z$: the $E_4$
term can be present in the anomaly  in $d=2$  just
for $z=2$.  The identification of the coefficient of the $E_4$ anomaly as
a monotonically decreasing quantity is possible only for $z=2$,
which forces $d$ to be even.
This justifies $d=2$ as the minimal choice
with non-trivial conformal anomaly.

\section{Newton-Cartan geometry}

Newton-Cartan (NC) geometry allows to reformulate Newtonian gravity
in a way which is coordinate independent; 
for a review see  \cite{gravitation}.
Recently, works by Son and collaborators \cite{Son:2005rv,Hoyos:2011ez,Son:2013rqa,Geracie:2014nka}
showed that it can be used as a powerful tool to study condensed matter systems
with galilean invariance; the main idea is to use it
 as source  for energy-momentum tensor for quantum field theory
 description of several condensed matter systems.
 NC geometry is an interesting topic by itself;
some useful references include \cite{Duval:1983pb,Duval:1984cj,Duval:2009vt,Brauner:2014jaa,
Jensen:2014aia,Jensen:2014wha,Bergshoeff:2014uea,Fuini:2015yva,Banerjee:2014pya,Bergshoeff:2015uaa}.
Strongly-coupled system with Galilean invariance can be studied 
holographically \cite{Son:2008ye,Balasubramanian:2008dm};
 also in this approach the NC geometry is a natural
formalism \cite{Christensen:2013lma,
Christensen:2013rfa,Andrade:2014iia,Hartong:2014oma,Hartong:2014pma}.

The basic data of Newton-Cartan  geometry
are a positive definite symmetric tensor $h^{\mu \nu}$  with rank $d$
(which corresponds to the spatial inverse metric)
 and a nowhere-vanishing vector $n_\mu$ (which corresponds to the local time direction),
with the condition 
\beq
n_\mu h^{\mu \a}=0  \, .
\eeq
The 1-form $n=n_\mu dx^\mu$ is not necessarily closed
(indeed it should not be if we want to use it as source for the energy current).
Another necessary ingredient is a non-dynamical gauge field $A_\mu$,
which acts as a source for the particle number symmetry.

In order to be able to define a metric with lower indices and a connection, one should 
also choose a velocity field $v^\mu$, with the condition $n_\mu v^\mu =1$.
Given $(h^{\a \b}, n_\mu, v^\nu)$ one can then uniquely define $h_{\mu \nu}$,
with:
\beq
h^{\mu \a}h_{\a \nu}=\delta^\mu_\nu -v^\mu n_\nu=P^\mu_\nu \, , \qquad
h_{\mu \a} v^\a =0 \, ,
 \eeq
 where $P^\mu_\nu$ is the projector onto spatial directions.
 
 Causality induces the following Frobenius integrability condition
 on the 1-form $n$:
 \beq 
 n \wedge dn =0, 
 \label{causale}
 \eeq
 see e.g. in \cite{Geracie:2014nka, Arav:2014goa}.
 This constraint is equivalent to the fact that $n$
 can be locally expressed as $n=g \, df$, where $f,g$ are functions.
The condition in eq.~(\ref{causale}) admits 
a non-zero torsion\footnote{Torsionless condition is equivalent to $dn=0$.}
 in the NC connection \cite{Geracie:2014nka}; this is chosen in
a way which is compatible with causality.
 
The symmetries of the Newton-Cartan theory include,
besides diffeomorphisms and local $U(1)$ gauge symmetry, 
the Milne boosts, which is a local version of the Galilean boost.
If we denote by $\psi_\mu$ the local parameter of the Milne boost,
the Newton-Cartan geometry fields transform in the following way:
\bea
v'^\mu & = & v^\mu+h^{\mu \nu} \psi_\nu \, \nl
h'_{\mu \nu} & = & h_{\mu \nu} -(n_\mu P_\nu^\rho+ n_\nu P_\mu^\rho) \psi_\rho
+n_\mu n_\nu h^{\rho \sigma} \psi_\rho \psi_\sigma \, , \nl
A'_\mu & = & A_\mu+P^\rho_\mu \psi_\rho -\frac{1}{2} n_\mu h^{\a \b} \psi_\a \psi_\b \, , 
\eea
while $n_\mu$ and $h^{\mu \nu}$ are invariant.
The following quantities are Milne invariants:
\beq
v_A^\a = v^\a - h^{\a \xi} A_\xi \, , \qquad
(h_A)_{\a \b}=h_{\a \b} +A_\a n_\b + A_\b n_\a \, , \qquad
\phi_A = A^2 - 2 v \cdot A \, .
\eeq
where $A^2=h^{\mu \nu}A_\mu A_\nu$
and $A \cdot v = v^\mu A_\mu$.

In the case in which a $2+1$ dimensional theory with 
non-zero magnetic momentum coupling is 
coupled to gravity, the Milne boost transformations of $A^\mu$ 
must be  appropriately modified, 
see e.g. \cite{Son:2013rqa,Jensen:2014aia}.
This situation is physically important for the quantum Hall effect.
In this case there is no parity invariance; we leave this 
as a topic for future investigations.

\subsection{Newton-Cartan theory and DLCQ}

It is possible to get non-relativistic Newton-Cartan theory
with zero magnetic momentum coupling as a dimensional 
reduction along the null direction $x^-$ of the following relativistic
non-degenerate metric \cite{Duval:1983pb,Duval:1984cj}:
\bea
\label{DLCQmetric}
G_{MN} & = & \left(
\begin{array}{cc}
 0& n_\mu \\ n_\nu  \, \, \, & n_\mu A_\nu + n_\nu A_\mu + h_{\mu \nu} \\
  \end{array}\right) =
  \left(
\begin{array}{cc}
 0 \, & n_\mu \\ n_\nu  \, &  (h_A)_{\mu \nu} \\
  \end{array}\right)\, ,   \nl
  G^{MN} & = &
  \left(
\begin{array}{cc}
 A^2-2 v \cdot A \, \, \, & v^\mu - h^{\mu \sigma} A_\sigma \\ 
v^\nu - h^{\nu \sigma} A_\sigma  \,\, \,&  h^{\mu \nu} \\
  \end{array}\right) =
    \left(
\begin{array}{cc}
\phi_A \, & v_A^\mu \\ 
v_A^\mu \, &  h^{\mu \nu} \\
  \end{array}\right) 
 \, .
\eea
The first row and column of this metric correspond to the null direction $x^-$,
the others to $x^\mu$. The metric components are independent from $x^-$, and
the following null vector generates an isometry:
\beq
n^M=(1,0, \dots) \, , \qquad n_{M}=(0,n_\mu) \, .
\eeq
We will refer to eq.~(\ref{DLCQmetric})
as the discrete light-cone quantization (DLCQ) metric.
The 1-form $n$ is naturally embedded in $d+2$ dimensions
as $n=n_A dx^A$.
Different ways to decompose $G_{\mu \nu}$ as a linear combination
of $h_{\mu \nu}$ and $A_{(\mu} n_{\nu)}$ are related by a Milne boost transformation.

The Levi-Civita connection from the metric in eq.~(\ref{DLCQmetric}) is as follows:
\bea
\label{levicivita}
& & \Gamma^{-}_{--}=\Gamma^\mu_{--}=0 \, , \qquad
\Gamma^-_{\alpha -}  =  \frac{1}{2} v_A^\s \tilde{F}_{\a \s}
 \, ,
\qquad
\Gamma^\mu_{\a -}= \frac{1}{2} h^{\mu \s} 
\tilde{F}_{\a \s}
\, ,
 \nl
& &\Gamma^-_{\a \b}= \frac{1}{2} \le \phi_A 
\tilde{S}_{\a \b}
+v_A^\s 
Q_{\a \b \s}
\ri \, ,
 \qquad 
\Gamma^\mu_{\a \b}= \frac{1}{2} \le v_A^\mu \tilde{S}_{\a \b}
+h^{\mu \s} Q_{\a \b \s}
  \ri \, ,
\eea
where we introduced the convenient quantities:
\bea
\tilde{F}_{\a \mu} & = &\p_\a n_\mu -\p_\mu n_\a  \, , \nl
\tilde{S}_{\a \b} & = &\p_\a n_\b + \p_\b n_\a \, ,   \nl
Q_{\a \b \sigma} & = & (\p_\a (h_A)_{\b \s} +\p_\b (h_A)_{\a \s} - \p_\s (h_A)_{\a \b}) \, . 
\eea
We should not confuse the
 DLCQ Levi-Civita connection in eq.~(\ref{levicivita})
with the NC connection, e.g. \cite{Geracie:2014nka}.
Indeed, the former is torsionless and the other
admits a non-vanishing torsion if $dn \neq 0$.

Let us denote by
$D_A$  the covariant derivative 
from the connection~(\ref{levicivita}).
Also we denote with $R$, $R_{ABCD}$ and $R_{AB}$
the  scalar curvature, the Riemann and the Ricci tensors 
associated with the Levi-Civita connection
of the metric eq.~(\ref{DLCQmetric}).

From the condition that $n_M$
is a Killing vector for the metric, it follows:
\beq
0=\li_n (G_{M N})=D_M n_N+D_N n_M \, .
\eeq
Using the fact that $n_A$ is null, one gets also
$D_M n^M=0$ and $n^S D_S n_M=0$.
We can define the 2-form 
\beq
\tilde{F}=\tilde{F}_{AB} dx^A \wedge dx^B =2 d n \, .
\eeq
In components:
\beq
\tilde{F}_{MN}=\p_M n_N -\p_N n_M=2 D_M n_N \, , \qquad \tilde{F}_{- \a} =0 \, .
\label{FTILDE}
\eeq

\section{Conformal anomaly ($z=2$)}

\subsection{Consequences of the causality constraint}

The causality condition eq.~(\ref{causale})
implies that $dn=n \wedge w$
for some one form $w$:
\beq 
\tilde{F}_{A B} = n_{[A} w_{B]} \, , \qquad w_A=(0, w_\a) \, ,
\qquad n^A w_A = 0 \, .
 \label{semp0}
\eeq
In fact, if $n \wedge dn=0$, 
it is locally possible to 
write 
\beq 
n=n_A dx^A= e^g \p_A f dx^A \, ,
\eeq
 where 
$g,f$ are two functions.
Then $\tilde{F}=2 e^g dg \wedge df$, 
and so a possible choice is 
\beq
w_A=-2 \p_A g \, .
\label{scelta}
\eeq
Note that $w_A$ in eq.~(\ref{semp0}) is not uniquely determined;
for example it could be shifted by
\beq
w_A \rightarrow w_A+p n_A \, ,
\label{shift}
\eeq
 where $p$ is an arbitrary function 
 (independent from $x^-$)
without affecting $\tilde{F}_{AB}$.
The vector $w_B$ determines also the covariant derivative of $n_A$ (see eq.~(\ref{FTILDE})).

\subsection{Classification of tensors}

In this section we will show
that scalars built with direct contractions of curvature tensors
with $n_A$ vanish; 
the only terms that we will need to keep track of
are the contractions with $w_A$, which is the
object that parameterize the derivative of $n_A$.

We define the following tensors:
\bea
 K^A_{\pa M}=R^A_{\pa B M N} n^B n^N \, , 
 & \qquad &
  K=K^A_{\pa A}=R_{AB} n^A n^B \, , \nl
T^A_{\pa BC}=R^A_{\pa BCD} n^D \, , 
& \qquad &
T_B=T^A_{\pa BA} \, .
\label{TTENSOR} 
\eea
Using eq.~(\ref{semp0}), we get that
\beq
K_{AB}= \chi \, n_A n_B \, , \qquad \chi=\frac{1}{16} G^{MN} w_M w_N\, .
 \label{semp1}
\eeq
Then just the $K^{--}$ component of $K^{AB}$ is non-zero; moreover
$K^A_{\pa B}$ is nihilpotent i.e. $K^A_{\pa B} K^B_{\pa C}=0$ and traceless $K=K^A_{\pa A}=0$.
Also note that $\chi$ is invariant under the shift
in eq.~(\ref{shift}).

Let us define
\beq
\Omega_{A B}=\frac{1}{16} (w_A w_B -4 D_A w_B) \, ,  \qquad
\Omega=\Omega_{AB} G^{AB} \, .
\label{omiga}
\eeq
We can use the ambiguity in eq.~(\ref{shift})
to render $\Omega_{AB}$ symmetric,
using the choice in eq.~(\ref{scelta}).
The following property is useful:
\beq
\Omega_{AB} n^B =\Omega_{B A} n^B= \chi \, n_A \, ,
\label{omiga2}
\eeq
We can write the tensors (\ref{TTENSOR})
in terms of $\Omega_{AB}$:
 \beq
T_{ABC}=\Omega_{CB} n_A - \Omega_{CA} n_B \, ,
 \qquad
 T_A=R_{AB} n^B=(\chi-\Omega) \, n_A  \, .
 \label{semp2}
\eeq

Due to the fact that $n^A$ generates an isometry,
the following Lie derivatives vanish:
\beq
\li_n (X) = 0 \, , \qquad
X=R_{MNPQ}\, , R_{MN}  \, , R \, ,
K_{AB}\, , T_A \, , T_{ABC} \, .
\label{lielie}
\eeq

Let us consider a generic term
contained in the anomaly.
One can set up a systematic procedure
to get rid of the terms which contain an explicit $n_A$
dependence:
\begin{itemize}
\item We can trade 
any derivative of $n_A$ with products of $n_A$, $w_A$
and derivatives of $w_A$, by repeated use of
eq.~(\ref{FTILDE}) and (\ref{semp0}).
\item  Every time $n^A$ is contracted with one of the tensors $X$, we can
 use eqs.~(\ref{semp1})-(\ref{semp2}) to simplify it to direct product
 of $n^A$, $X$, $w^A$ and its derivatives.
\item When $n^A$ is contracted with a derivative
 $D_A X$, we can use eq.~(\ref{lielie}) to trade it 
 with some other direct product of $n^A$, $w^A$ and $X$,
using eq.~(\ref{FTILDE}) and (\ref{semp0}).
 \item When $n^B$ is contracted with a tensor inside a derivative,
 we can use Leibniz rule:
 \beq
 n^B D_A (X_{B})=   D_A (n^B X_{B})-  D_A (n^B) X_{B} \, ,
 \eeq
 to reduce to a contraction with a tensor with does not have 
 an explicit derivative.
\end{itemize}
 At the end of this procedure, we will end up just with terms in which $n^A$ is only contracted
 with $n_A$ or $w_A$, which both vanish.
 For example:
 \beq
 T^{ABC} T_{ABC}=-2 \Omega^{CA} n^B \Omega_{CB} n_A = -2 \chi^2 n^A n_A =0 \, .
 \eeq

 This argument shows that in order to classify non-vanishing terms in the anomaly,
 it is sufficient to consider only terms which can be written using
 the curvatures, $w^A$ and their derivatives. It is not necessary
 to consider terms which depend
 explicitly on
 $n^A$ and its derivatives because they vanish.
This will be used to show that only a finite number of
terms survives out of
 of the infinite family
considered in \cite{Jensen:2014hqa}.

\subsection{Weyl transformations}

In our notation,  the metric transforms as follows under Weyl transformations:
\beq
n_\mu \arr e^{2 \s} n_\mu \, , \qquad
h_{\mu \nu} \arr e^{2 \s} h_{\mu \nu} \qquad
v^\mu \arr e^{-2 \s} v^\mu \, , \qquad
h^{\mu \nu} \arr e^{-2 \s} h^{\mu \nu} \, ,
\label{weyl1}
\eeq
while the coordinates do not transform.

A Weyl transformation on the Newton-Cartan background
is equivalent to a  Weyl transformation in the extra-dimensional metric
in eq.~(\ref{DLCQmetric})  
which is independent from the $x^-$ coordinate:
\beq
n^A D_A \s =0 \, .
\eeq
Eq.~(\ref{weyl1}) fixes the transformation properties of the metric in eq.~(\ref{DLCQmetric}):
\beq
G_{MN} \arr e^{2 \s} G_{MN} \, , 
\qquad
G^{MN} \arr e^{-2 \s} G^{MN} \, ,
\qquad
n^A \arr n^A \, , \qquad 
n_A \arr  e^{2 \s} n_A \, .
\label{weyl2}
\eeq
Moreover, starting from of eq.~(\ref{weyl2}) it is possible to derive
the Weyl transformation of all the building blocks entering the anomaly,
see table \ref{tabella}.

\begin{table}[h]     
\begin{center}    
\begin{tabular}  {|l|l|} \hline Term & Weyl variation \\
\hline
$\Gamma^A_{BC}$ & $\delta^A_B D_C \s +\delta^A_C D_B \s 
-G_{BC} D^A \s$ \\
$ D^2 \phi $ & $D^2 (\delta \phi)
 -2 \s D^2 \phi +2 D^A \s D_A \phi$ \\
 $D_A V_B$ & $D_A (\delta V_B) - V_A D_B \s -V_B D_A \s + G_{AB} V^M D_M \s$ \\
$R^A_{\pa BMN}$ & $- \delta^A_M D_B D_N \s +\delta^A_N D_B D_M \s
+ G_{BM} D^A D_N \s   - G_{BN} D^A D_M \s $ \\
$R$ & $-2 \s R - 6 D^2 \s$  \\
$\chi$ & $-2 \s \chi-\frac{1}{2} w^A \p_A \s$  \\
$\Omega$ & $-2 \s \Omega -w^A \p_A \s +D^2 \s$  \\
$w_A$ & $ -4 D_A \s$\\
$R_{M N}$ & $-2 D_M D_N \s - G_{M N} D^2 \s $ \\ 
$ \Omega_{CB}$ & $D_B D_C \s-\frac{1}{4} G_{BC} w^K D_K \s$ \\
\hline
\end{tabular}   
\caption{\footnotesize Weyl variation of the basic fields entering the anomaly.
Here $V_B$ is a generic vector and $\phi$ a generic scalar.}
\label{tabella}  
\end{center}
\end{table}

The NC measure $\sqrt{\det \gamma}$, where
\beq
\gamma_{\mu \nu}= n_\mu n_\nu + h_{\mu \nu} \, ,
\eeq
is also equal to $\sqrt{- \det {G_{AB}}}=\sqrt{-G}$.
The Weyl weight of this measure is equal to $d+2$
in $d$ spatial dimensions. Consequently, in odd spatial
dimensions it is not possible to obtain a trace anomaly.
We can write anomalies only for even spatial dimension.
In the case of $d=0$ all the curvatures and tensor
such as $w^A$ are zero and the anomaly
is vanishing. We specialize to the minimal dimension $d=2$
in which the anomaly survives.

\subsection{The anomaly}

The anomaly can be written
as a linear combination of $16$ linearly-independent 
terms, built from DLCQ tensors\footnote{NC $U(1)$ gauge  transformations are included in the DLCQ diffeomorphisms. Consequently, any DLCQ scalar is automatically
$U(1)$ gauge invariant. Milne boosts invariance is also automatic as it is related to
equivalent ways to decompose $G_{\mu \nu}$ as a linear combination
of $h_{\mu \nu}$ and $A_{(\mu}n_{\nu )}$.
}: 
\beq
\mathcal{A}_\s =\sum_{k=1}^{16} b_k \mathcal{A}_\s^k \, ,
\qquad
 \mathcal{A}_\s^k =\int \sqrt{-G}  \, d^{4}x \, \left(\s \, A_k \right) \, ,
\eeq
where $b_k$ are anomaly coefficients and 
\bea
A_1= E_4 \, , \qquad  A_2=W^2 \, , & \qquad &  A_3=D^2 R \, , 
\qquad   A_4=R^2  \, ,    \nl
A_5=\chi^2 \, , \qquad A_6=\Omega^2 \, , &\qquad&  A_7=\chi \Omega  \, , \qquad
A_8=\chi R \, , \nl
 A_9=\Omega R \, , \qquad
A_{10}=\Omega_{AB} \Omega^{AB} \, , & \qquad & 
 A_{11}=\Omega_{AB} R^{AB}  \, , \qquad
A_{12}= \Omega_{AB} w^A w^B \, ,   \nl  
A_{13}= R_{AB} w^A w^B \, , 
\qquad
A_{14} = w^A D_A R \, , & \qquad & 
A_{15}=D^2 \chi \, , \qquad
A_{16}=D^2 \Omega \, . 
\label{BAE}
\eea
In our conventions the
 Euler density $E_4$ and
 the square of Weyl tensor $W^2$ are:
\beq
E_4=R^2_{ABMN} -4 R^2_{AB} +R^2 \, ,
\qquad
W^2_{ABMN}=R^2_{ABMN} -2 R^2_{AB} +\frac{1}{3} R^2 \, .
\eeq
This basis cover the most generic term with the
correct Weyl weight; all the other terms 
with the same Weyl weight can be reduced to these
 ones using integration by parts.

One may wonder about the explicit form of $E_4$
in term of $2+1$ dimensional fields, at least in some simple example.
In the special case of flat space metric
$h_{\mu \nu}= {\rm Diag} \, (0,1,1)$ and generic $n_\a$
consistent with the condition (\ref{causale}),
  \beq
 E_4=8 (6 \chi^2 -4 \chi \Omega + \Omega^2 - \Omega_{AB} \Omega^{AB}) \, .
  \eeq
This expression is not valid if a non-zero spatial curvature is present.
A general expression for $E_4$ is given in eq.~(\ref{E4esplicito}).

\subsection{Consistency conditions}

We solve the Wess-Zumino consistency conditions 
for the Weyl symmetry \cite{Bonora:1983ff}:
\beq
\Delta^{WZ}_{\s_1 \s_2} \mathcal{A} =\delta^W_{\s_2} \mathcal{A}_{\s_1}-
\delta^W_{\s_1} \mathcal{A}_{\s_2} = 0 \, .
\eeq
Using integrations by parts and formulas in appendix \ref{AppeB},
the commutator of the two Weyl variations
can be written as a linear combination
of several independent expressions $C_k$:
\beq
\Delta^{WZ}_{\s_1 \s_2} \mathcal{A}^k=
\int  \sqrt{-G} \, d^4 x \, \left( \sum_{m=1}^8 M^{km} C_m \right) \, ,
\qquad k=1  \ldots 16 \, ,
\eeq
where 
\bea
&& C_1=\commuA D^A R \, , \qquad
 C_2=\commuA D^A \chi  \, , \qquad
C_3=\commuA D^A \Omega \, , \nl
&& C_4=\commuA R w^A  , \qquad
 C_5=\commuA \chi w^A  \, , \qquad 
C_6=\commuA \Omega w^A \, , \nl
&& C_7=\commuC w^A  \, , \qquad C_8= \commuA D^2 w^A \,   .
\label{BACH}
\eea

The transpose of the matrix $M^{km}$ is:
\beq
(M^t)^{mk}=\left(
\begin{array}{cccccccccccccccc}
 0 & 0 & 0 & 12 & 0 & 0 & 0 & 0 & -1 & 0 & -\frac{1}{2} & 0 & 0 & -4 & 0 & 0 \\
 0 & 0 & 0 & 0 & 0 & 0 & -1 & 6 & 0 & -1 & 1 & 8 & 64 & 0 & 2 & 4 \\
 0 & 0 & 0 & 0 & 0 & -2 & 0 & 0 & 6 & 0 & 1 & 0 & -32 & 0 & -2 & -4 \\
 0 & 0 & 0 & 0 & 0 & 0 & 0 & -\frac{1}{2} & -1 & 0 & -\frac{1}{4} & 0 & 0 & -2 & 0 & 0 \\
 0 & 0 & 0 & 0 & -1 & 0 & -1 & 0 & 0 & -\frac{1}{2} & \frac{1}{2} & -16 & 8 & 0 & 0 & 0 \\
 0 & 0 & 0 & 0 & 0 & -2 & -\frac{1}{2} & 0 & 0 & 0 & -\frac{1}{2} & 4 & -8 & 0 & 0 & 0 \\
 0 & 0 & 0 & 0 & 0 & 0 & 0 & 0 & 0 & 0 & 0 & 0 & 0 & -6 & -\frac{1}{2} & -1 \\
  0 & 0 & 0 & 0 & 0 & 0 & 0 & 0 & 0 & \frac{1}{2} & -\frac{1}{2} & 0 & -8 & 0 & 0 & 0 
\end{array}
\right) \, .
\label{BACH2}
\eeq 
This matrix has $9$ null eigenvectors, which correspond to the consistent 
combination of the various anomaly terms.

On the other hand, not every consistent combination corresponds to a 
genuine anomaly. In fact we must eliminate terms which are the Weyl variation of local counterterms:
\beq
\delta^W \left( \sum_{k=1}^{16}  \int \sqrt{-G} \, d^4 x \, \s b_k A_k \right)
\eeq
because they correspond to scheme-dependent terms.
 These local counterterms are chosen in such a way
that they are invariant under all the NC symmetries: diffeomorphisms, gauge $U(1)$ 
and Milne boosts.
This eliminates $6$ out of $9$ consistent terms.

\section{Conclusions}

For a $d=2$ Schr\"odinger invariant theory coupled to a NC background,
the three genuinely independent terms in the conformal 
anomaly can be chosen as:
\beq
\mathcal{A}= a \s E_4 + c \s W^2 + b  \s J^2  +\mathcal{A}_{ct} \, ,
  \label{risultato}
\eeq
where
\beq
 J  =\Omega-2 \chi +\frac{R}{6}\, .
  \label{jj}
 \eeq
Both $J^2$ and $W^2$ have vanishing Weyl variation and
they correspond to type B anomalies \cite{Deser:1993yx}. 

The scheme-dependent part $\mathcal{A}_{ct}$ is an arbitrary linear combination
of the following terms:
\bea
\label{schemedependent} 
&& \s D^2 R \, , \qquad \s D^2 (\Omega-2 \chi) \, , \qquad
\s \le 12 \chi^2-4 \chi \Omega-\frac{1}{2} \Omega_{A B} w^A w^B \ri\, ,
\nl
&& \s \le 2 R \chi -2 R \Omega +\frac{w^A D_A R}{2} -6 D^2 \chi \ri \, , \nl
&& \s \le
2 \chi R -2 \Omega R +4 \Omega_{AB} R^{AB} -\frac{R_{AB} w^A w^B}{4}
\ri \, , \nl
&& \s \le -\Omega^2+2 \chi \Omega+\Omega_{AB}^2 -\frac{\Omega_{AB} w^A w^B}{8}
+\frac{R_{AB} w^A w^B}{16} \ri \, . 
\eea
This result is rather different from the Lifshitz case
\cite{Adam:2009gq,Baggio:2011ha,Griffin:2011xs,Arav:2014goa}
where just type B anomalies are present.
The analog of the $E_4$ term in this case can be reabsorbed
by a local counterterm.
 This is due to the fact that the number of admissible counterterms is bigger
in the Lifshitz case, because the symmetry content is smaller.
It is surprising that the structure of the anomaly in eq.~(\ref{risultato}) 
 is similar to the one in the relativistic case with an extra spatial dimension.
The only difference is that an extra type B anomaly $J^2$ is present, 
and there are several more scheme-depedent terms. 
In both cases, the $E_4$ term is the only anomaly 
with non-trivial Weyl variation.
For general $z$, the Weyl weights do not allow the presence
of the $E_4$ term in the anomaly (see appendix \ref{AppeA}).

The coefficient $a$ is a natural candidate for a decreasing quantity 
from the UV to IR conformal fixed point.
It will be possible to check this statement in concrete
examples, both at weak and at strong coupling.
It would be interesting to clarify the relation 
between the anomaly and  Anti-de Sitter radius 
appearing in the holographic calculation in
\cite{Myers:2010tj} and \cite{Liu:2015xxa}.
Another exciting direction is to explore the relation between
scale invariance and conformal invariance in the non relativistic case
\cite{Nakayama:2009ww}.

We did not consider the case of non-zero magnetic momentum coupling
$g_s$
where a modified Milne boost transformation is required.
In this case, the  DLCQ description \cite{Duval:1983pb,Duval:1984cj}
 cannot be used.
As a consequence, the classification of the possible terms entering the anomaly looks harder. Nonetheless, the $E_4$ term written as a function of NC quantities is still Milne boost invariant
(see eq.~(\ref{E4esplicito}))
 because it does not contain
$A_\mu$.
Therefore, even if $g_s\ne 0$, the $a$-coefficient in eq.~(\ref{risultato})
is still a candidate for a monotonically decreasing quantity.

\section*{Acknowledgments}

We are grateful  to Carlos Hoyos and Mauro Spera for useful discussions.

\section{Erratum}
Due to an overcounting of the number of linearly-independent
terms in the basis, the type A anomaly disappears 
because it can be eliminated
by local counterterms $\mathcal{A}_{\rm ct}$.
The complete anomaly then is:
\beq
2 T^0_0+T^i_i=\mathcal{A}= b  \s J^2  + \mathcal{A}_{\rm ct} \, .
\eeq 
This is in agreement with \cite{Arav:2016xjc}.

The basis of the anomaly in eq.~(\ref{BAE}) turns out to be redundant,
due to the presence of the following relations
(which are valid if the Frobenius condition $n \wedge dn=0$ holds):
\bea
&& W^2=12 J^2 \, , \nl
&&
E_4=72 \chi^2 -4 \chi R -48 \chi \Omega + 8 \Omega^2 -8 \Omega_{AB} \Omega^{AB} \, , \nl
&&
(R_{AB} +2 \Omega_{AB}) w^A w^B = 8 \chi(R-6 \chi +4 \Omega) \, ,
\nl
&&
 (R_{AB} +2 \Omega_{AB}) \Omega^{AB}
= 12 \chi^2 +\frac{1}{2} \Omega (R+4 \Omega) - \chi (R + 9 \Omega) \, .
\label{4equazioniimportanti}
\eea
The derivation of these equations was explained to us by
Igal Arav and Shira Chapman, to whom we are deeply grateful.

Consequently, the elements $A_1$, $A_2$, $A_{11}$ and $A_{13}$
 can be written as linear combinations of the remaining terms.
The basis of commutators of two Weyl variations in eq.~(\ref{BACH})
is also redundant:
\beq
C_8 = 8 C_2 -4 C_3 +\frac{1}{2} C_4 -5 C_5 +2 C_6 \, .
\eeq
Then, the matrix in eq.~(\ref{BACH2}) becomes 
\beq
(M^t)^{mk}=\left(
\begin{array}{cccccccccccc}
 0 & 12 & 0 & 0 & 0 & 0 & -1 & 0 & 0 & -4 & 0 & 0 \\
 0 & 0 & 0 & 0 & -1 & 6 & 0 & 3 & 8 & 0 & 2 & 4 \\
 0 & 0 & 0 & -2 & 0 & 0 & 6 & -2 & 0 & 0 & -2 & -4 \\
 0 & 0 & 0 & 0 & 0 & -\frac{1}{2} & -1 & \frac{1}{4} & 0 & -2 & 0 & 0 \\
 0 & 0 & -1 & 0 & -1 & 0 & 0 & -3 & -16 & 0 & 0 & 0 \\
 0 & 0 & 0 & -2 & -\frac{1}{2} & 0 & 0 & 1 & 4 & 0 & 0 & 0 \\
 0 & 0 & 0 & 0 & 0 & 0 & 0 & 0 & 0 & -6 & -\frac{1}{2} & -1 \\
\end{array}
\right) \, .
\eeq 
The null space of this matrix (which has dimension $6$)
can be written as direct sum of the $5$ counterterms:
\bea
\label{schemedependent2} 
&& \s D^2 R \, , \qquad \s D^2 (\Omega-2 \chi) \, , \qquad
\s \le 12 \chi^2-4 \chi \Omega-\frac{1}{2} \Omega_{A B} w^A w^B \ri\, ,
\nl
&& \s \le 2 R \chi -2 R \Omega +\frac{w^A D_A R}{2} -6 D^2 \chi \ri \, , \nl
&&
\s \le-9 \chi^2 -\Omega^2+6 \chi \Omega+\Omega_{AB}^2 
+\frac{ \chi R}{2} \ri
 \, , 
\eea
and the remaining type B anomaly $J^2$.

We are grateful to  Igal Arav and Shira Chapman for useful discussions.

\section*{Appendix}
\addtocontents{toc}{\protect\setcounter{tocdepth}{1}}
\appendix

\section{Considerations on the case with general dynamical exponent $z \neq 2$}

\label{AppeA}

In the case of general dynamical exponent $z$,
the Weyl transformations in eq.~(\ref{weyl1})
are modified into:
\beq
n_\mu \arr e^{z \s} n_\mu \, , \qquad
v^\mu \arr e^{-z \s} v^\mu \, , \qquad
A_\mu \arr e^{(2-z)\s} A_\mu \, , \qquad
\sqrt{\det \gamma}  \arr e^{(d+z) \s}\sqrt{\det \gamma} \, ,
\label{weylbis}
\eeq
while the ones for $h^{\a \b}$, $h_{\a \b}$
are unchanged.
Since the DLCQ metric components in eq.~(\ref{DLCQmetric})
 do not transform homogeneously,
one may wonder if the Weyl weights of the quadratic extra-dimensional 
curvature invariants are still well-defined.

Given a contravariant vector $V^A$, 
there is the following relation between the Weyl weight of the $V^\a$
and of the $V^-$ components:
\beq
V^\a \arr e^{x \s} V^\a \, , \qquad
V^- \arr e^{(x+2-z) \s} V^- \, .
\label{rule1}
\eeq
Conversely, for covariant vectors:
\beq
W_\a \arr e^{y \s} W_\a \, , \qquad
W_- \arr e^{(y-2+z)\s} W_- \, .
\label{rule2}
\eeq
Note that  the 
weight of the $V^- W_-$ contraction is the same as  
the one of $V^\a W_\a$ and does not depend on $z$.
The same rules can be iterated for tensors, for example:
\beq
G^{\a \b} \arr e^{-2 \s} G^{\a \b} \, , \qquad
G^{\a -} \arr e^{-z \s} G^{\a -} \, , \qquad
G^{- -} \arr e^{(2-2z) \s} G^{--} \, .
\eeq
Also, these rules work for the vectors $w^A$ and $n^A$
and are consistent with the lowering and raising index procedure.
Using the Levi-Civita connection eq.~(\ref{levicivita}), it can be shown
that if these rule are valid for a tensor, they are valid also for its derivatives.
Using the definition of the Riemann tensor in term of commutator of covariant derivatives,
it follows that they apply also to the Riemann tensor.

This shows that for every tensor that we can build from the curvatures,
$n^A$, $w^A$ and their derivatives, the rules in eqs.~(\ref{rule1})
and (\ref{rule2}) can be used to determine the weight
of some components in which some minus index is present in term of
the components with greek indices.
Moreover, every scalar defined contracting these tensors
has a definite weight, which can be determined
looking at the weight of the part where just
greek indices are contracted.
This shows that the Weyl weight of terms built from curvature
and $w_A$ and its derivatives are well-defined and independent of $z$.

Looking at table \ref{tabella},
all the ingredients that can enter the anomaly 
have even Weyl weight; then, taking
the weight of the measure $\sqrt{\gamma}$
into account, a non-zero trace anomaly exists just for
\beq
z=2n-d \, ,
\eeq
where $n$ is a positive integer.

\section{Useful formulas}

\label{AppeB}

The following identities are useful in manipulating 
the anomaly and its Weyl variation:
\beq
D_A w^A =4 (\chi-\Omega) \, , \qquad
D_A w_B =\frac{1}{4} w_A w_B -4 \Omega_{AB} \, ,
\eeq
\beq
\Omega^{AB} w_B =w^A \chi-2 D^A \chi \, ,
 \qquad 
 w^A D_A \chi  =
8 \chi^2 -\frac{1}{2} \Omega_{AB} w^A w^B \, ,
\eeq
\beq
16 D^2 \chi=
2\le\frac{w_A w_B}{4} -4 \Omega_{AB}\ri^2  +2 (D^2 w_A) w^A  \, ,
\eeq
\beq
(D^2 w_A) w^A =
8 D^2 \chi -16 \chi^2 -16 \Omega_{AB}^2 +2 \Omega_{AB} w^A w^B \, ,
\eeq
\beq	
w^B D^A \Omega_{AB}=
12 \chi^2 -4 \chi \Omega -2 D^2 \chi -\frac{3}{4} \Omega_{AB} w^A w^B
+4 \Omega_{AB}^2 \, ,
\label{formu}
\eeq
\bea
D^A D^B \Omega_{AB}
&=& \Omega^2-2 \chi \Omega -\Omega_{AB}^2
+ D^2 \Omega + R^{AB} \Omega_{AB} \nl
 &+&\frac{1}{8} \le \Omega_{AB} w^A w^B
- R_{AB} w^A w^B  - w^A D_A R  \ri \, ,
\eea
\beq
w^A D_A \Omega=
12 \chi^2+ 4 \Omega_{AB}^2 - \Omega_{AB} w^A w^B -2 D^2 \chi +\frac{1}{4} R_{AB} w^A w^B \, ,
\eeq
\beq
D^2 D_A \s =D_A D^2 \s +R_{BA} D^B \s \, ,
\eeq
\beq
D^2  w^A =D_A (D_B w^B) + R_{AB} w^B \, .
\label{batistuta}
\eeq

An explicit expression for $E_4$ in terms of NC fields is:
\bea
\label{E4esplicito}
E_4&=&8 (\Omega-2 \chi)^2
-8h^{\alpha \mu} h^{\beta \nu} \Omega_{\alpha \beta} \Omega_{\mu \nu}
+ h^{\alpha \mu} h^{\beta \nu} h^{\rho \xi} h^{\sigma \eta} \tilde{R}_{\alpha \rho \beta \sigma}
\tilde{R}_{\mu \xi \nu \eta} 
-4 h^{\alpha \mu} h^{\beta \nu} h^{\rho \sigma}
 h^{\xi \eta} \tilde{R}_{\alpha \rho \beta \sigma}
 \tilde{R}_{\mu \xi \nu \eta} 
\nl
&&
+16  h^{\alpha \mu} h^{\beta \nu} h^{\rho \sigma} \tilde{R}_{\alpha \rho \beta \sigma}   \Omega_{\mu \nu}
 +( \tilde{R}_{\alpha \rho \beta \sigma} h^{\a \b} h^{\rho \sigma})^2
 +2 (6 \chi - 4 \Omega) \tilde{R}_{\alpha \rho \beta \sigma} h^{\a \b} h^{\rho \sigma}
\eea
where
\bea
\tilde{R}_{\alpha \rho \beta \sigma} &=&
-\frac{1}{2} \partial_{\alpha} \partial_{\beta} h_{\rho \sigma} 
- \frac{1}{2} \partial_{\rho} \partial_{\sigma} h_{\alpha \beta}
+ \frac{1}{2} \partial_{\alpha} \partial_{\sigma} h_{\rho \beta}
+ \frac{1}{2} \partial_{\rho} \partial_{\beta} h_{\alpha \sigma} \nl
&&
- \frac{1}{4} h^{\tau_{1} \tau_{2}}(H_{\alpha \beta \tau_{1}} H_{\rho \sigma \tau_{2}} 
- H_{\alpha \sigma \tau_{1}} H_{\rho \beta \tau_{2}}) 
 \nl
&& 
- \frac{1}{4} v^{\tau} (\tilde{S}_{\alpha \beta} H_{\rho \sigma \tau} + \tilde{S}_{\rho \sigma} H_{\alpha \beta \tau}
- \tilde{S}_{\alpha \sigma} H_{\rho \beta \tau} -
 \tilde{S}_{\rho \beta}  H_{\alpha \sigma \tau}) \, .
\eea
and
\beq
H_{\a \b \sigma}= (\p_\a (h)_{\b \s} +\p_\b (h)_{\a \s} - \p_\s (h)_{\a \b}) \, .
\eeq



\begin{thebibliography}{99}
 
 
\bibitem{Arav:2016xjc}
  I.~Arav, S.~Chapman and Y.~Oz,
  arXiv:1601.06795 [hep-th].
 
\bibitem{Zamolodchikov:1986gt}
  A.~B.~Zamolodchikov,
  JETP Lett.\  {\bf 43} (1986) 730
   [Pisma Zh.\ Eksp.\ Teor.\ Fiz.\  {\bf 43} (1986) 565].
 

 
 \bibitem{Cardy:1988cwa}
  J.~L.~Cardy,
  Phys.\ Lett.\ B {\bf 215} (1988) 749.
 
\bibitem{Bonora:1985cq}
  L.~Bonora, P.~Pasti and M.~Bregola,
  Class.\ Quant.\ Grav.\  {\bf 3} (1986) 635.

\bibitem{Duff:1977ay}
  M.~J.~Duff,
  Nucl.\ Phys.\ B {\bf 125} (1977) 334.

\bibitem{Duff:1993wm}
  M.~J.~Duff,
  Class.\ Quant.\ Grav.\  {\bf 11} (1994) 1387
  [hep-th/9308075].
 

 
 \bibitem{Osborn:1989td}
  H.~Osborn,
  Phys.\ Lett.\ B {\bf 222} (1989) 97.
 
 \bibitem{Jack:1990eb}
  I.~Jack and H.~Osborn,
  Nucl.\ Phys.\ B {\bf 343} (1990) 647.
 
 \bibitem{Osborn:1991gm}
  H.~Osborn,
  Nucl.\ Phys.\ B {\bf 363} (1991) 486.
  
  
  
\bibitem{Komargodski:2011vj}
  Z.~Komargodski and A.~Schwimmer,
  JHEP {\bf 1112} (2011) 099
  [arXiv:1107.3987 [hep-th]].
 
 \bibitem{Komargodski:2011xv}
  Z.~Komargodski,
  JHEP {\bf 1207} (2012) 069
  [arXiv:1112.4538 [hep-th]].
  
\bibitem{Myers:2010tj}
  R.~C.~Myers and A.~Sinha,
  JHEP {\bf 1101} (2011) 125
  [arXiv:1011.5819 [hep-th]].
  
  \bibitem{Jafferis:2011zi}
  D.~L.~Jafferis, I.~R.~Klebanov, S.~S.~Pufu and B.~R.~Safdi,
  JHEP {\bf 1106} (2011) 102
  [arXiv:1103.1181 [hep-th]].
  
  
 

\bibitem{Adam:2009gq}
  I.~Adam, I.~V.~Melnikov and S.~Theisen,
  JHEP {\bf 0909} (2009) 130
  [arXiv:0907.2156 [hep-th]].

\bibitem{Baggio:2011ha}
  M.~Baggio, J.~de Boer and K.~Holsheimer,
  JHEP {\bf 1207} (2012) 099
  [arXiv:1112.6416 [hep-th]].

\bibitem{Griffin:2011xs}
  T.~Griffin, P.~Horava and C.~M.~Melby-Thompson,
  JHEP {\bf 1205} (2012) 010
  [arXiv:1112.5660 [hep-th]].

\bibitem{Arav:2014goa}
  I.~Arav, S.~Chapman and Y.~Oz,
  JHEP {\bf 1502} (2015) 078
  [arXiv:1410.5831 [hep-th]].
 
  
 
\bibitem{Deser:1993yx}
  S.~Deser and A.~Schwimmer,
  Phys.\ Lett.\ B {\bf 309} (1993) 279
  [hep-th/9302047].
  
  
 
  
\bibitem{Jensen:2014hqa}
  K.~Jensen,
  arXiv:1412.7750 [hep-th].
  
\bibitem{Bergman:1993kq}
  O.~Bergman and G.~Lozano,
  Annals Phys.\  {\bf 229} (1994) 416
  doi:10.1006/aphy.1994.1013
  [hep-th/9302116].

\bibitem{Nishida:2007pj}
  Y.~Nishida and D.~T.~Son,
  Phys.\ Rev.\ D {\bf 76} (2007) 086004
  doi:10.1103/PhysRevD.76.086004
  [arXiv:0706.3746 [hep-th]].
  
 
\bibitem{Son:2008ye}
  D.~T.~Son,
  Phys.\ Rev.\ D {\bf 78} (2008) 046003
  doi:10.1103/PhysRevD.78.046003
  [arXiv:0804.3972 [hep-th]].


\bibitem{Balasubramanian:2008dm}
  K.~Balasubramanian and J.~McGreevy,
  Phys.\ Rev.\ Lett.\  {\bf 101} (2008) 061601
  doi:10.1103/PhysRevLett.101.061601
  [arXiv:0804.4053 [hep-th]].
  
\bibitem{Duval:1983pb}
  C.~Duval and H.~P.~Kunzle,
  Gen.\ Rel.\ Grav.\  {\bf 16} (1984) 333.

\bibitem{Duval:1984cj}
  C.~Duval, G.~Burdet, H.~P.~Kunzle and M.~Perrin,
  Phys.\ Rev.\ D {\bf 31} (1985) 1841.




\bibitem{gravitation}
 Charles W. Misner, Kip S. Thorne and John Archibald Wheeler (1973), Gravitation,
 San Francisco: W. H. Freeman, ISBN 978-0-7167-0344-0.



\bibitem{Son:2005rv}
  D.~T.~Son and M.~Wingate,
  Annals Phys.\  {\bf 321} (2006) 197
  [cond-mat/0509786].

\bibitem{Hoyos:2011ez}
  C.~Hoyos and D.~T.~Son,
  Phys.\ Rev.\ Lett.\  {\bf 108} (2012) 066805
  [arXiv:1109.2651 [cond-mat.mes-hall]].

\bibitem{Son:2013rqa}
  D.~T.~Son,
  arXiv:1306.0638 [cond-mat.mes-hall].

\bibitem{Geracie:2014nka}
  M.~Geracie, D.~T.~Son, C.~Wu and S.~F.~Wu,
  Phys.\ Rev.\ D {\bf 91} (2015) 045030
  [arXiv:1407.1252 [cond-mat.mes-hall]].

\bibitem{Brauner:2014jaa}
  T.~Brauner, S.~Endlich, A.~Monin and R.~Penco,
  Phys.\ Rev.\ D {\bf 90} (2014) 10,  105016
  [arXiv:1407.7730 [hep-th]].
  
\bibitem{Jensen:2014aia}
  K.~Jensen,
  arXiv:1408.6855 [hep-th].

\bibitem{Jensen:2014wha}
  K.~Jensen and A.~Karch,
  JHEP {\bf 1504} (2015) 155
  [arXiv:1412.2738 [hep-th]].


 
\bibitem{Bergshoeff:2014uea}
  E.~A.~Bergshoeff, J.~Hartong and J.~Rosseel,
  Class.\ Quant.\ Grav.\  {\bf 32} (2015) 13,  135017
  [arXiv:1409.5555 [hep-th]].
  
\bibitem{Fuini:2015yva}
  J.~F.~Fuini, A.~Karch and C.~F.~Uhlemann,
  arXiv:1510.03852 [hep-th].
  


  
\bibitem{Duval:2009vt}
  C.~Duval and P.~A.~Horvathy,
  J.\ Phys.\ A {\bf 42} (2009) 465206
  doi:10.1088/1751-8113/42/46/465206
  [arXiv:0904.0531 [math-ph]].

\bibitem{Banerjee:2014pya}
  R.~Banerjee, A.~Mitra and P.~Mukherjee,
  Phys.\ Lett.\ B {\bf 737} (2014) 369
  doi:10.1016/j.physletb.2014.09.004
  [arXiv:1404.4491 [gr-qc]].
  
\bibitem{Bergshoeff:2015uaa}
  E.~Bergshoeff, J.~Rosseel and T.~Zojer,
  Class.\ Quant.\ Grav.\  {\bf 32} (2015) 20,  205003
  doi:10.1088/0264-9381/32/20/205003
  [arXiv:1505.02095 [hep-th]].




\bibitem{Christensen:2013lma}
  M.~H.~Christensen, J.~Hartong, N.~A.~Obers and B.~Rollier,
  Phys.\ Rev.\ D {\bf 89} (2014) 061901
  doi:10.1103/PhysRevD.89.061901
  [arXiv:1311.4794 [hep-th]].

\bibitem{Christensen:2013rfa}
  M.~H.~Christensen, J.~Hartong, N.~A.~Obers and B.~Rollier,
  JHEP {\bf 1401} (2014) 057
  doi:10.1007/JHEP01(2014)057
  [arXiv:1311.6471 [hep-th]].

\bibitem{Andrade:2014iia}
  T.~Andrade, C.~Keeler, A.~Peach and S.~F.~Ross,
  Class.\ Quant.\ Grav.\  {\bf 32} (2015) 3,  035015
  doi:10.1088/0264-9381/32/3/035015
  [arXiv:1408.7103 [hep-th]].

\bibitem{Hartong:2014oma}
  J.~Hartong, E.~Kiritsis and N.~A.~Obers,
  Phys.\ Lett.\ B {\bf 746} (2015) 318
  doi:10.1016/j.physletb.2015.05.010
  [arXiv:1409.1519 [hep-th]].



\bibitem{Hartong:2014pma}
  J.~Hartong, E.~Kiritsis and N.~A.~Obers,
  Phys.\ Rev.\ D {\bf 92} (2015) 066003
  [arXiv:1409.1522 [hep-th]].
  
  
  
\bibitem{Bonora:1983ff}
  L.~Bonora, P.~Cotta-Ramusino and C.~Reina,
  Phys.\ Lett.\ B {\bf 126} (1983) 305.

\bibitem{Liu:2015xxa}
  J.~T.~Liu and W.~Zhong,
  arXiv:1510.06975 [hep-th].

\bibitem{Nakayama:2009ww}
  Y.~Nakayama,
  Int.\ J.\ Mod.\ Phys.\ A {\bf 24} (2009) 6197
  [arXiv:0906.4112 [hep-th]].





 \end{thebibliography}
\end{document}